\documentclass[twocolumn,prd,aps,showpacs,amsmath,amssymb]{revtex4-1}
\usepackage{graphics}
\usepackage{graphicx}
\usepackage{dcolumn}
\usepackage{bm}
\usepackage{mathrsfs}
\usepackage{pstricks}
\usepackage{color}
\usepackage{slashed}
\usepackage{amsmath}
\usepackage{epsfig}
\usepackage{amsfonts}
\usepackage{amssymb}
\newcommand{\Fstar}{\raisebox{.2ex}{$\stackrel{*}{F}$}{}}

\def\beq{\begin{equation}}
\def\eeq{\end{equation}}
\def\bea{\begin{eqnarray}}
\def\eea{\end{eqnarray}}

\begin{document}

\title{Multirefringence phenomena in nonlinear electrodynamics}
\author{Vitorio A. De Lorenci}
\email{delorenci@unifei.edu.br}
\affiliation{Instituto de F\'{\i}sica e Qu\'{\i}mica, Universidade
Federal de Itajub\'a, Itajub\'a, MG 37500-903, Brazil}
\affiliation{Institute of Cosmology, Department of Physics and Astronomy,
Tufts University, Medford, Massachusetts 02155, USA}
\author{Renato Klippert}
\email{klippert@unifei.edu.br}
\affiliation{Instituto de Matem\'atica e Computa\c{c}\~ao, Universidade Federal de Itajub\'a, Itajub\'a, MG 37500-903,
Brazil}
\author{Shi-Yuan Li}
\email{lishy@sdu.edu.cn}
\affiliation{School of Physics, Shandong University, Jinan, 250100, P. R. China}
\author{Jonas P. Pereira}
 \email{jonaspedro.pereira@icranet.org}
\affiliation{Universit\'e de Nice Sophia Antipolis, 06103 Nice Cedex 2, France}
\affiliation{Dipartamento di Fisica and ICRA, Sapienza Universit\`a di Roma, I-00185 Rome, Italy}

\begin{abstract}
Wave propagation in nonlinear theories of the electromagnetism described by Lagrangian densities
dependent upon its two local invariants $L(F, G)$ is revisited. On the light of the recent findings in
metamaterials, it is here shown that trirefringence is also a possible phenomenon to occur in the
realm of such nonlinear theories. A specific model exhibiting this effect is investigated both in terms
of phase and group velocities. It is claimed that wave propagation in
some well known nonlinear models for spin-one fields, like QED and QCD in certain regimes, may
exhibit trirefringence.
\end{abstract}

\textit{}

\maketitle

\section{Introduction}
\label{introduction}
As it is well known, nonlinear theories of electromagnetism
exhibit birefringence phenomenon. The most popular example
appears in the quantum electrodynamics (QED) where polarization
effects are activated in the limit of large fields
($B_{\scriptscriptstyle cr} \sim E_{\scriptscriptstyle cr} = m^2c^3/e\hbar$),
inducing an effective optical axis in the vacuum. In such situation,
a light ray is expected to split in two rays propagating with
different velocities \cite{Birula,Adler,delorenci2000a,delorenci2000b}.
An experimental setup designed to measure the birefringent properties of the
QED vacuum was long ago proposed \cite{Iacoppini}. However, direct measurements
of this effect are not yet conclusive and are still under consideration \cite{valle2010}.
The influence of the nontrivial vacua on the propagation of electromagnetic waves
was discussed in several distinct physical configurations
\cite{Latorre,Drumond,Shore,Scharnhorst,Barton,Dittrich}.
Conditions for the occurrence of birefringence of gluon fields was also
studied \cite{shi2008}.

In the context of material media, birefringence effects are expect to occur
in several distinct situations. It occurs naturally, for instance in
certain crystals presenting optical axes \cite{Landau, born}, or artificially when optical axes
are induced by means of external applied electromagnetic fields \cite{delorenci2008,delorenci2004}.
Nowadays, birefringent materials and methods including this effect have been
incorporated in several technological
devices \cite{paschotta2008}. Birefringence is also a powerful optical tool to investigate
properties of new materials, biological systems and others \cite{bio,astro1,astro2}.

On the other hand, trirefringence was only recently considered
as a possible phenomenon in material media. It was measured \cite{tri1}
in tailored photonic crystals \cite{Joannopoulos}, and the theoretical
description of this effect in media characterized  by effective
dielectric coefficients was proposed \cite{delorenci2012}.
In this case \cite{delorenci2012}, only when some of the dielectric
coefficients are negative, could trirefringence take place.
Metamaterials \cite{smith2,smith,veselago,shelby} seem to be good candidates
for supporting this effect, due to the controllability of their dielectric tensors.
With the present day technology of producing such new media, it is expected that
trirefringence will play some important role in technology of optical
systems, as birefringence has done.

Usually, effects occurring in the realm of Maxwell electromagnetism
in material media are expected to occur in the realm of nonlinear electromagnetic
theories. It is possible to build up analogue models between these two
domains where the coefficients describing a specific dielectric
medium are mapped as derivatives of the Lagrangian density describing a
nonlinear theory. In this way, trirefringence should also be a possible
effect in nonlinear electromagnetism.
By deriving and using the general description for wave propagation in
the limit of geometrical optics, in this paper we show that trirefringence is
in fact a possible effect in nonlinear electrodynamics.
It is not our purpose to set the general conditions a model
must fulfill in order to present trirefringence, but only to
show the effect as a possible one in the domain of nonlinear electromagnetism.
A particular model is thus investigated where such phenomenon is shown to occur provided
that convenient external fields are set. The phase and group velocities of
the waves are derived, as well as the corresponding polarization vectors.
A numerical example is graphically studied.

The model examined in the paper corresponds to the effective Lagrangian density
for QED in the regime of large fields. The trirefringence phenomenon is shown to occur whenever
the model applies, although its measurability requires the control of very large fields. A possible
arena to search for this effect could be the special fluids recently produced by high energy collisions,
as addressed later in the concluding section. Further, this model is also useful in the context
of analogue models in material media, where the predicted effect could be tested on optical systems,
for instance in metamaterials.

In the next section, nonlinear electromagnetism is briefly revisited
and the field equations are presented in terms of the general
two parameters Lagrangian density.
In Sec. \ref{wavepropagation}, the corresponding wave propagation is examined.
The eigenvalue problem is stated and solved,
resulting in the general fourth degree equation for the phase velocities.
This equation is solved for a specific nonlinear model in Sec. \ref{trirefringence}.
The corresponding polarization states and the description of the effect in terms
of group velocities are also discussed.
Conclusions and final remarks are presented in Sec. \ref{conclusion}.

Throughout this paper we employ the Minkowski metric $\eta_{\mu\nu} = {\rm diag} (+1,-1,-1,-1)$.
The completely skew-symmetric tensor $\eta^{\alpha\beta\mu\nu}$ is defined
by $\eta^{0123} = 1$. We set the units such that the velocity of light in empty space
is $c=1$.

\section{Nonlinear electrodynamics: field equations}
\label{fieldequations}
Nonlinear Abelian theories for electromagnetism can be formulated by means of the
general Lagrangian density $L=L(F,G)$, where $F$ and $G$ are the two local gauge invariants
of the electromagnetic field. These invariants are defined in terms of the
electromagnetic tensor field $F_{\mu\nu}$, and its dual
\begin{equation}
\label{dual}
\Fstar_{\alpha\beta} = \frac{1}{2}\eta_{\alpha\beta}{}^{\sigma\tau}F_{\sigma\tau},
\end{equation}
as
\begin{eqnarray}
\label{F}
F&=&F^{\mu\nu}F_{\mu\nu}\\
\label{G}
G&=&F^{\mu\nu}\Fstar_{\mu\nu}.
\end{eqnarray}
In terms of the electric $\vec{E}$ and magnetic $\vec{B}$ field strengths we have
$F=-2(E^2-B^2)$ and $G=-4\vec{E}\cdot\vec{B}$.

The field equation can be obtained from the least action principle
and it can be presented as \cite{delorenci2000a},
\begin{equation}
\label{field1}
2N^{\mu\nu\alpha\beta}F_{\alpha\beta ,\nu}
+ L_F F^{\mu\nu}{}_{,\nu} = 0,
\end{equation}
where $N^{\mu\nu\alpha\beta}$ is defined by
\begin{eqnarray}
N^{\mu\nu\alpha\beta} &\doteq&
L_{FF}F^{\mu\nu}F^{\alpha\beta}
+ L_{GG} \Fstar^{\mu\nu}\Fstar^{\alpha\beta}
\nonumber \\
&&+ L_{FG}\left(
F^{\mu\nu}\Fstar^{\alpha\beta} +
\Fstar^{\mu\nu}F^{\alpha\beta}\right).
\label{21}
\end{eqnarray}
Use is being made here of the notation
$L_{X^{1}X^{2}\cdots X^{n}}=\partial^n L/\partial X^{1}\partial X^{2}\cdots\partial X^{n}$
previously introduced \cite{delorenci2000a},
where each $X^i$ is one of the two invariants $F$ or $G$ upon which the Lagrangian $L$ arbitrarily depends.
We notice that the above defined rank-4 tensor presents the following symmetries:
$N^{\mu\nu\alpha\beta} = -N^{\nu\mu\alpha\beta}$, $N^{\mu\nu\alpha\beta} =-N^{\mu\nu\beta\alpha}$ and
$N^{\mu\nu\alpha\beta} = N^{\alpha\beta\mu\nu}$.

In addition to Eq. (\ref{field1}), $F_{\mu\nu}$ satisfies the Bianchi identity
$\Fstar^{\mu\nu}{}_{,\nu}=0$, which implies in the existence of a potential
vector $A_\mu$ as
\begin{equation}
F_{\mu\nu} = A_{\mu,\nu}-A_{\nu,\mu}.
\label{potential}
\end{equation}

\section{Nonlinear electrodynamics: wave propagation}
\label{wavepropagation}
Let us now discuss the propagation of electromagnetic waves in the general formulation
of nonlinear electrodynamics. We restrict ourselves to the propagation of monochromatic
waves in the limit imposed by geometrical optics \cite{Landau,born}. The method of
field discontinuities will be used, which can be briefly stated as follows
\cite{Hadamard,delorenci2000a}.

Consider a differentiable inextendible oriented borderless hypersurface $\Sigma$,
defined locally by $\phi(x^\mu)=0$, where $\phi$ is a real differentiable scalar field
which locally is a function of the spacetime coordinates $x^\mu=(t,\vec x)$.
Let $U^{{}+{}}$ be the spacetime points whose coordinates satisfy $\phi(x^\mu)>0$,
and similarly $U^{{}-{}}$ be such that $\phi(x^\mu)<0$.
Let $P$ be any given point of $\Sigma$.
For each sufficiently small $r>0$, let $V_r(P)$ be a neighborhood of $P$
which consists of the spacetime points $Q$ whose Euclidean distance from $P$
is $[(t_Q-t_P)^2+||\vec x{}_Q-\vec x{}_P||^2]^{(1/2)}$ smaller than $r$.
Let $P^{{}+{}}\in U^{{}+{}}\cap V_r(P)$ and $P^{{}-{}}\in U^{{}-{}}\cap V_r(P)$
be any two neighbor points from $P$ arbitrarily chosen at opposite sides of $\Sigma$.
Let $f$ be any given tensor field defined at $V_r(P)$.
The Hadamard discontinuity at $P$ of $f$ across $\Sigma$ is defined as
\begin{equation}
[f]_\Sigma(P)\doteq\lim_{r\rightarrow0^{{}+{}}}\left[f(P^{{}+{}})-f(P^{{}-{}})\right]
\label{discontinuity}.
\end{equation}

Suppose $f$ such that $[f]_\Sigma=0$ for each $P\in\Sigma$.
Following Hadamard \cite{Hadamard}, we have
$[f_{,\lambda}]_\Sigma(P)=k_\lambda\bar f(P)$,
where $k_\lambda=\phi_{,\lambda}|_P$ is the normal vector to $\Sigma$ at $P$
and $\bar f$ is a tensor field defined at $\Sigma$
with the same rank and the same algebraic symmetries as those of $f$.

We assume the electromagnetic tensor field $F_{\mu\nu}$ to be smooth in each $U^\pm$,
but merely continuous at $\Sigma$
(that is to say, the $F_{\mu\nu}$ are continuous functions at $\Sigma$
but their derivatives may present discontinuities at $\Sigma$).
The Hadamard discontinuities at $\Sigma$ of Eq.~(\ref{field1}) and of the derivative
of Eq.~(\ref{potential}) lead to \cite{delorenci2000a}
\begin{equation}
\label{field1b}
f_{\beta\lambda} k^{\lambda} + \frac{2}{L_F}N_{\beta}{}^{\mu\nu\rho}f_{\nu\rho}k_{\mu} = 0
\end{equation}
and
\begin{equation}
f_{\alpha\beta} = \epsilon_\alpha k_\beta - \epsilon_\beta k_\alpha,
\label{field2b}
\end{equation}
where the quantities $f_{\alpha\beta}$
are related to the derivatives of $F_{\alpha\beta}$ on $\Sigma$
by $[F_{\alpha\beta,\lambda}]_\Sigma=f_{\alpha\beta}k_\lambda$ and $\epsilon_\mu$ is
the polarization vector $[A_{\mu,\alpha\beta}]_{{}_{\Sigma}} = e_\mu k_\alpha k_\beta$.
We set $k_{\lambda} = \omega V_\lambda+q_\lambda$ as the wave 4-vector,
where $V_\lambda=\delta^0_\lambda$ is the 4-velocity of the observer
which decomposes $F^{\mu\nu}$ into electric and magnetic fields.
The components of this 4-vector $k_\lambda$ are thus the frequency $\omega$
and the wave vector $\vec{q} = q\hat{q}$, where $\|\hat{q}\|^2=-\hat{q}^\lambda \hat{q}_\lambda =1$.

Taking together
Eqs. (\ref{field1b}) and (\ref{field2b}),
we obtain the general eigenvalue equation \cite{Birula,delorenci2000a}
\begin{equation}
Z^\mu{}_\nu \epsilon^\nu = 0,
\label{Ze}
\end{equation}
where
\begin{equation}
\label{Z}
Z^\mu{}_\nu \doteq k^2\delta^\mu{}_\nu
+\frac{4}{L_F}N^{\mu\alpha}{}_{\nu\beta}k_\alpha k^\beta,
\end{equation}
with $k^2 \doteq k^\lambda k_\lambda$.
Nontrivial solutions of Eq. (\ref{Ze}) can be found only if
$\det \mid Z_{\mu\nu}\mid \,=\, 0$,
the well known generalized Fresnel equation, and yields
\begin{equation}
\alpha(k^2)^2 + \beta f^2 k^2 + \gamma(f^2)^2 = 0,
\label{1}
\end{equation}
where $f^2 \doteq F^{\alpha\mu}F_{\alpha}{}^\nu k_\mu k_\nu$, and
\begin{eqnarray}
\alpha &=& L_F^2+2L_F(GL_{FG}-FL_{GG})
\nonumber\\
&&-(L_{FF}L_{GG}-L_{FG}^2)G^2,
\label{2} \\
\beta &=& 4L_F(L_{FF}+L_{GG})-8(L_{FF}L_{GG}-L_{FG}^2)F,
\label{3} \\
\gamma &=& 16(L_{FF}L_{GG}-L_{FG}^2).
\label{4}
\end{eqnarray}
One can also recast the quantity $f^2$ that appears in Eq. (\ref{1}) as
\begin{equation}
f^2 = (\vec{q}\cdot\vec{E})^2 - \omega^2E^2+ (\hat{q}\cdot\vec{B})^2-q^2B^2
+ 2\omega\vec{q}\cdot\vec{E}\times\vec{B}.
\label{20}
\end{equation}

The phase velocity $v\doteq \omega/q$
of the electromagnetic waves can be obtained from Eq. (\ref{1}).
In fact, it is straightforward to show that this equation
can be presented as a fourth-degree equation for the phase velocity $v$ as
\begin{equation}
a_4 v^4 + a_3 v^3 + a_2 v^2 + a_1 v + a_0 = 0,
\label{26}
\end{equation}
where we have defined
\begin{eqnarray}
a_4 &\doteq& \alpha - \beta E^2 + \gamma E^4
\label{a4} \\
a_3 &\doteq&  2(\beta - 2\gamma E^2)\hat{q}\cdot\vec{E}\times\vec{B}
\label{a3} \\
a_2 &\doteq& -2\alpha + \beta [E^2 - B^2 + (\hat{q}\cdot\vec{E})^2
+ (\hat{q}\cdot\vec{B})^2]
\nonumber\\
&&+ 2\gamma\{2(\hat{q}\cdot\vec{E}\times\vec{B})^2
-[(\hat{q}\cdot\vec{E})^2 +(\hat{q}\cdot\vec{B})^2
\nonumber \\
&& - B^2]E^2\}
\label{a2} \\
a_1 &\doteq& -2\{\beta -2\gamma[(\hat{q}\cdot\vec{E})^2 +(\hat{q}\cdot\vec{B})^2
\nonumber\\
&&- B^2]\}
\hat{q}\cdot\vec{E}\times\vec{B}
\label{a1} \\
a_0 &\doteq& \alpha +\beta[B^2 - (\hat{q}\cdot\vec{E})^2 - (\hat{q}\cdot\vec{B})^2]
\nonumber \\
&&+\gamma[B^2 - (\hat{q}\cdot\vec{E})^2 - (\hat{q}\cdot\vec{B})^2]^2.
\label{a0}
\end{eqnarray}
As stated by Eq. (\ref{26}), we can find up to four solutions for the phase velocity
in the same wave direction. In the next section we will analyze some special
cases where multirefringence phenomena may occur.

Dispersion relations for light propagation in nonlinear electrodynamics can also be
investigated by means of the photon mass operator \cite{tsai1974,tsai1974b}.
In such context the propagation of photons in homogeneous magnetic field was investigated
long ago \cite{tsai1976} and birefringence phenomena was described for some field configurations.

\section{A model for trirefringence}
\label{trirefringence}
In what follows we shall study a particular model for nonlinear electromagnetism
which presents interesting multirefringence features. Let the one-parameter nonlinear
Lagrangian density
\begin{eqnarray}
L_{NL} = - \frac{1}{4}b_0 F \log\frac{F}{\lambda^2},
\label{LYM}
\end{eqnarray}
where $b_0$ and $\lambda$ are constants. Particularly these constants can be chosen
in order to split the above model in a Maxwellian part plus a nonlinear
contribution.

This model appears in different contexts in the literature,
for instance as the effective Lagrangian
density for quantum electrodynamics (QED) \cite{schwinger1951} in the regime of
large fields \cite{weisskopf,schwinger1954a,schwinger1954b,elmfors1993}.
In this case the constants in Eq. (\ref{LYM}) are related to the fine
structure constant $\alpha$ and to the critical
electromagnetic fields for which vacuum polarization effects begin to
become important.
Furthermore, with an appropriate choice of dielectric coefficients this model can describe several kinds of magnetic materials which can be used
as analogue systems to investigate properties of the vacuum of the non-Abelian
gauge field \cite{pagels1978}.
A more detailed discussion about this issue is presented in the concluding section.

\subsection{Phase velocity}
The study of wave propagation in this special case can be done by
following the lines presented in Sec. \ref{wavepropagation}.
We will investigate multirefringence phenomena using $L_{NL}$ in the Abelian case
only.

To study a simplified case, let us assume constant external
electric $\vec{E}=E\hat{x}$ and magnetic $\vec{B}=B\hat{y}$ fields,
much larger than their wave counterparts.
The wave vector $\vec{q}$ is assumed to lie in
the $xz$-plane so that, $\hat{q}\cdot\hat{y} = 0$, $\hat{q}\cdot\hat{x} = \sin\theta$,
and $\hat{q}\cdot\hat{z} = \cos\theta$. Thus, $\theta$
is the angle between $\hat{q}$ and $\hat{z}$ directions.
From the above notation, $\hat{q}\cdot\vec{E}\times\vec{B}=EB\cos\theta$.
Taking $L_{NL}$ in Eq. (\ref{26}) we obtain the following results
\begin{eqnarray}
v_0 &=& 1,
\label{v0}\\
v_\pm &=&\frac{2EB\cos\theta\pm\sqrt{(\chi-2B^2)(\chi-2E^2\cos^2\theta)}}{\chi}
\label{vpm},
\end{eqnarray}
where it was introduced the shortcut
\begin{equation}
\chi\doteq2E^2-\frac{F}{2}\left(1+\log\frac{F}{\lambda^2}\right)
\label{chi}.
\end{equation}
The quantity $v_0$ is isotropic and does not depend on any
choice for the configuration of the fields nor direction of propagation.
However, the other solutions $v_\pm$ will be different for different
configurations of fields and $\hat{q}$ direction, set by the angle $\theta$.
As one can see, three different velocities in the same direction
occur provided the square-root is smaller than $2BE\cos\theta$. This naturally imposes conditions on the fields.
We observe that trirefringence will occur in a region defined by $-\theta_c<\theta<\theta_c$, where
\begin{equation}
\theta_c=\arccos \left\{\sqrt{-\frac{F}{4E^2}\left(3+\log\frac{F}{\lambda^2}\right)}\right\}
\label{tr1}.
\end{equation}
Therefore, it does exist iff
\begin{equation}
0<F<F_c,\;\;\; \frac{F_c}{\lambda^2}\doteq e^{-3}\simeq 0.0498\label{tr2}
\end{equation}
and
\begin{equation}
E>E_m,\;\;\; E_m^2\doteq -\frac{F}{4}\left(3+\log\frac{F}{\lambda^2}\right)\label{trm}.
\end{equation}
Birefringence takes place in the regions $\theta_c<\theta<\pi-\theta_c$ and
$-(\pi -\theta_c)<\theta<-\theta_c$. For the remaining angles, just the
ordinary wave exists. When $F=F_c$, trirefringence is not present for any direction.
If one keeps $F$ fixed and satisfying Eq. (\ref{tr2}) and decreases the electric field
[satisfying Eq. (\ref{trm})], then the region where trirefringence takes place decreases and
the difference between the moduli of the extraordinary solutions increases.
\begin{figure}[!hbt]
\leavevmode
\centering
\includegraphics[scale = .85]{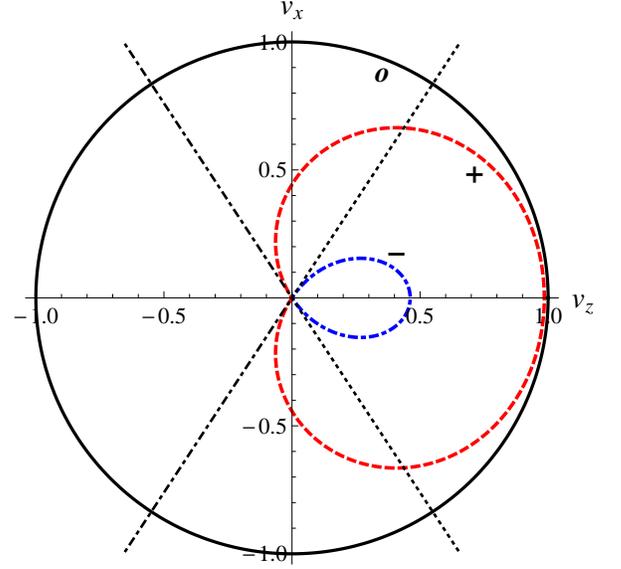}
\caption{{\small\sf (color online).
Normal surfaces for the nonlinear theory described by Eq. (\ref{LYM}).
The ordinary wave is represented by the circular thick line and the extraordinary waves
are represented by the dashed and dot-dashed curves. The symbols $+$, and $-$ and $o$ indicate the solutions presented in
Eqs. (\ref{vpm}) and (\ref{v0}).
The chosen values of the electric and magnetic fields are $B=0.1\lambda$ and $E=0.09\lambda$.
These values satisfy  Eqs. (\ref{tr2}) and (\ref{trm}); hence trirefringence is present.
Since this is the case, also birefringence and one refraction must take place. The regions where these effects take place are related
to the critical angle, $\theta_c$, as given by Eq. (\ref{tr1}). The trirefringent region lies between the dotted straight lines, while the
birefringent regions are constituted by the angles limited by the dotted and dot-dashed straight lines. The region presenting one
refraction lies between the dot-dashed straight lines.}}
\label{fig0}
\end{figure}
In Fig. \ref{fig0}, the normal surfaces \cite{Landau,born} associated with Eqs. (\ref{v0}) and (\ref{vpm})
are plotted for a selected set of the fields.
It can easily be seen there the region exhibiting trirefringence, as anticipated, for fields satisfying
Eqs. (\ref{tr2}) and (\ref{trm}).

\subsection{Polarization}
\label{polarization}
Let us now return to the Fresnel-like eigenvalue Eq. (\ref{Ze}).
The matrix $Z^\mu{}_\nu$ given by Eq. (\ref{Z}) reduces, for the Lagrangian in Eq. (\ref{LYM}), to
\begin{eqnarray}
\frac{Z^\mu{}_\nu}{q^2}&=&(v^2-1)\delta^\mu_\nu+\frac{4}{F\left(1+\log\frac{F}{\lambda^2}\right)}
\left[v E^\mu+(\hat q\cdot\vec E)V^\mu\right.
\nonumber\\
&&\left.+(\hat q\times\vec B)^\mu\right]
\left[v E_\nu+(\hat q\cdot\vec E)V_\nu+(\hat q\times\vec B)_\nu\right]
\label{ZLYM}.
\end{eqnarray}
This suggests that the polarization vector $\epsilon^\nu$ should conveniently be decomposed
as a linear combination of the three vectors which appear in $Z^\mu{}_\nu$ as
\begin{equation}
\epsilon^\nu=a v E^\nu + b (\hat q\cdot\vec E)V^\nu + c (\hat q\times\vec B)^\nu + d k^\nu
\label{epsilon},
\end{equation}
where $a,b,c$ are arbitrary constants with the same physical dimension.
The fourth term, with an arbitrary constant $d$,
was introduced because Eq. (\ref{field2b}) remains unchanged by it.
Equation (\ref{Ze}) then reads
\begin{eqnarray}
[a(v^2-1)-Y]v E^\mu+[b(v^2-1)-Y](\hat q\cdot\vec E)V^\mu
\nonumber\\
+[c(v^2-1)-Y](\hat q\times\vec B)^\mu=0
\label{eigenvalue},
\end{eqnarray}
where $Y$ is a shortcut for
\begin{eqnarray}
Y&\doteq&\frac{4}{F\left(1+\log\frac{F}{\lambda^2}\right)}\left[av^2E^2-(a+c)v\hat{q}\cdot(\vec E\times\vec B)\right.
\nonumber \\
&&\left.+c(\hat q\times\vec B)^2-b(\hat q\cdot\vec E)^2\right]
\label{X}.
\end{eqnarray}

For the $v^2=1$ case, Eq. (\ref{eigenvalue}) yields $Y=0$,
from which the polarization state is given by
$a=(\hat q\cdot\vec E)^2,\;b=E^2-(\hat q\times\vec B)^2,\;c=-(\hat q\cdot\vec E)^2$
up to a global multiplicative factor.
For the $v^2\neq1$ case, Eq. (\ref{eigenvalue}) yields $a=b=c=1$
up to a global multiplicative factor.
Assuming $\vec E\cdot\vec B=0$ and $\hat q\cdot\vec B=0$,
then the phase velocities for this case are given by Eq. (\ref{vpm}).
Once the physical parameters $a,b,c$ were found in either case,
then the particular gauge choice $d=-b(\hat q\cdot\vec E)/(qv)$
ensures $\epsilon^\nu$ to lie in the space orthogonal to $V^\nu$.

\subsection{Group velocity}
\label{groupvelocities}
\begin{figure}[!hbt]
\leavevmode
\centering
\includegraphics[scale = .85]{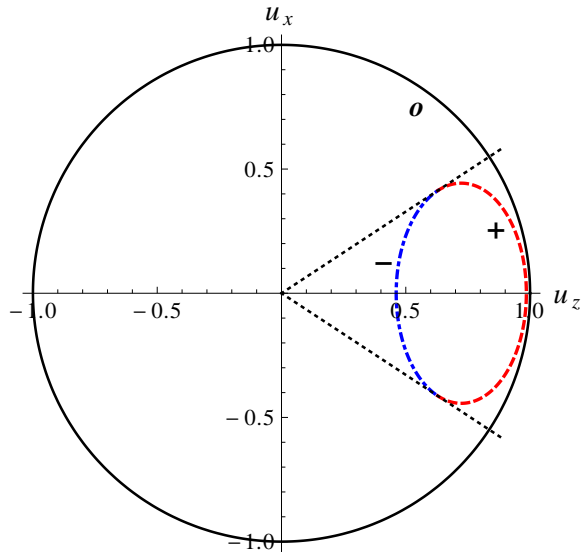}
\caption{{\small\sf (color online).
Ray velocities for the nonlinear model described by Eq. (\ref{LYM}).
The fields are the same as in Fig. \ref{fig0}.
The symbols $+$, $-$ and $o$ indicate the solutions presented in
Eqs. (\ref{vpm}) and (\ref{v0}) when substituted in Eqs. (\ref{gym1})--(\ref{gym3}). As it can be seen,
trirefringence occurs in the region lying between the dotted straight lines. In the complementary
region only the isotropic ray solution propagates.}}
\label{fig2}
\end{figure}

It is well known that, in the geometrical optics approximation the wave equation
is a linear equation for the perturbed fields, even in the context
of nonlinear electrodynamics \cite{Birula}. In this case wave
packets can be build up by superposing plane wave solutions, whose phase velocities
were obtained above. Thus, it is important to deal with the group velocities
$\vec u\doteq{\rm d}\omega/{\rm d}\vec q$ for the propagation analysis \cite{Landau}.
For the particular case of the nonlinear Lagrangian density $L_{NL}$,
with the same configuration of fields and wave vectors we assumed above,
the associated group velocities are
\begin{equation}
\vec{u}= u_x\hat{x}+u_z\hat{z}\label{gym1},
\end{equation}
where $u_x=\sin\theta$ and $u_z=\cos\theta$ for the ordinary $v=1$ mode,
thus stating that the group and phase velocities coincide for this case.
For the extraordinary modes, we obtain
\begin{equation}
u_x^\pm=\frac{\chi v_\pm^2-4EBv_\pm\,\cos\theta+2E^2\cos^2\theta}{\chi v_\pm-2EB\,\cos\theta}\,\sin\theta
\label{gym2},
\end{equation}
and
\begin{equation}
u_z^\pm=\frac{\chi v_\pm^2\,\cos\theta-2EBv_\pm\cos2\theta-2E^2\cos\theta\,\sin^2\theta}{\chi v_\pm-2EB\,\cos\theta}
\label{gym3},
\end{equation}
where we are considering $v_\pm$ as given by Eq. (\ref{vpm}).
Hence, two extraordinary rays and one ordinary ray can be found in the $xz$- plane.
But their dependence on direction must be investigated more carefully, as we shall do in what follows.
If one defines $\varphi$ as the angle between the group velocity and the $z$-axis,
then $\varphi=\theta$ for the $v=1$ case.
For the extraordinary modes, it follows from Eqs. (\ref{gym2})--(\ref{gym3}) that
\begin{equation}
\tan\varphi=\frac{(\chi v_\pm^2-4EBv_\pm\cos\theta+2E^2\cos^2\theta)\sin\theta}
{\chi v_\pm^2\,\cos\theta-2EBv_\pm\cos2\theta-2E^2\cos\theta\,\sin^2\theta}
\label{gym4},
\end{equation}
which gives us $\varphi$ as a function of $\theta$.
For the case of the extraordinary solutions, the analytical inversion of Eq. (\ref{gym4})
to give $\theta$ as a function of $\varphi$ is very involved.
Hence, numerical analyses turns out to be more clarifying.
Fig. \ref{fig2} summarizes such a numerical analysis for the same parameters assumed in Fig. \ref{fig0}.
One notices that the region of the $xz$ plane where the ordinary and extraordinary group velocities can be found 
constitutes a trirefringent region.
For the complementary region of the plane of propagation, there exists just the ordinary group velocity.

Comparing Figs. \ref{fig0} and \ref{fig2} we find angular sectors
for which plane waves associated with the extraordinary polarization
modes are supposed to propagate, but with no propagation of the
corresponding wave packets. This feature relies on the fact that, in our analysis, the wave packets for the extraordinary modes do propagate along
directions which are not generally equal to the directions of the
corresponding plane waves components, as it is explicitly shown in
Eq. (\ref{gym4}). This is due to the fact that the phase velocity is dependent upon the direction of the wave vector;
hence $\omega=v(\hat{q})q$. This dependence leads to a term in the group velocity that is perpendicular to the phase velocity.
The magnitude of such a term may be comparable to the magnitude of the phase velocity itself, yielding therefore to a
possibly different behavior of the two aforementioned velocities.

\section{Conclusion and Discussion}
\label{conclusion}
Trirefringence phenomena is not an effect exclusively occurring in nonlinear
metamaterials \cite{tri1,delorenci2012}. As shown here, it is possible to
formulate a nonlinear model describing electromagnetism where this effect is
also expected to occur.
Possible extensions of the presented model can be sought by adding the dual
invariant $G$, or else trying other nonlinear Lagrangian models.

As it is well known, in the regime of small fields QED is governed by the Euler-Heisenberg
effective Lagrangian density \cite{schwinger1951}. In this situation birefringence
effect is predicted to occur \cite{Birula}. Experiments are still under consideration
in order to confirm such prediction \cite{battesti2008,valle2010}.
However, when the regime of large fields is considered, the effective Lagrangian
governing QED presents the same form as the nonlinear model stated in Eq. (\ref{LYM}),
leading to the conclusion that trirefringence phenomenon is expected to occur
in this regime.

Multirefringence phenomena could also be found for systems described by
nonlinear Lagrangian densities which depend on non-Abelian gauge fields.
In such cases, the field strength tensors would not be gauge-invariant.
Nevertheless, it can be easily shown that the propagation of the field disturbances
would be described by the same equations presented above,
since only second-order derivatives of the gauge vector field
may present non-zero Hadamard discontinuities.

The nonlinear model discussed in the text
was proposed long ago \cite{savvidy1977,pagels1978,nielsen1979}
as the effective Lagrangian of quantum-chromodynamics (QCD)
or other Yang-Mills theories with non-trivial vacuum properties.
When considered for this purpose,
the Lorentz invariant parameter $F$ in  $L(F)$ is extrapolated to
be $F^{\mu \nu (a)}F_{\mu \nu (a)}$, where the index $(a)$ runs in the inner Non-Abelian group
space. Considering the regime of small coupling and taking the limit of
large mean fields ($F/\lambda^2 \gg 1$) a Lagrangian density
with the same functional form as appearing in Eq. (\ref{LYM}) is obtained \cite{pagels1978}.
In this context $b_0$ may be identified as a $\beta$-function coefficient
at leading order and $\lambda$ a constant related to the mass scale.
It was claimed \cite{adler1981} that the form for
$L_{NL}$ may also result in the leading terms in the limit of weak fields
($F/\lambda^2 \ll 1$), as in both cases $|\log(F/\lambda^2)| \gg 1$.
Hence trirefringence of non-Abelian gauge fields can also occur and in principle it can be
observed, provided the required field configuration is approached.
The practical arena for such kind of study is a quark-gluon plasma (QGP), recently observed
in high-energy heavy ion collision experiments where the gluon field is deconfined
and can propagate in the bulk of the QGP. Some symmetric field configuration has been
investigated, and a possible observable has been proposed \cite{shi2008}. With the progress of
measurements on asymmetries in the experiments, which is now a hot topic in RHIC and
LHC, other kind of field configurations can be further investigated.

\acknowledgments
This work was partially supported by the Brazilian CAPES (under scholarship BEX 18011/12-8), CNPq, FAPEMIG
and Chinese NSFC, NSFSC (Natural Science Foundation of Shandong Province of China)
research agencies. J.P.P. acknowledges the support given by the
Erasmus Mundus Joint Doctorate Program, under the Grant No. 2011-1640
from EACEA of the European Commission.



\end{document}